\documentclass[aps,twocolumn,showpacs,byrevtex]{revtex4-1}

\usepackage{graphicx}
\usepackage{times}
\usepackage{amsmath}
\usepackage{mathtools}
\usepackage{cancel}
\usepackage[T1]{fontenc}
\usepackage{color}
\usepackage{float}
\usepackage{subfigure}
\usepackage{lineno} 
\usepackage[colorlinks,linkcolor=blue,anchorcolor=blue,citecolor=blue]{hyperref}

\newcommand{\Ks}{K_{S}^{0}}

\newcommand{\pip}{\pi^{+}}
\newcommand{\pim}{\pi^{-}}
\newcommand{\pimp}{\pi^{\mp}}
\newcommand{\kpm}{K^{\pm}}

\newcommand{\kp}{K^{+}}
\newcommand{\km}{K^{-}}
\newcommand{\x}{X(3872)}

\newcommand{\ee}{e^+e^-}
\newcommand{\jpsi}{J/\psi}

\begin{document}

\title{\boldmath Search for the decays $X(3872)\to K_{S}^{0}K^{\pm}\pi^{\mp}$ and $K^*(892)\bar{K}$ at BESIII}

    \author{
M.~Ablikim$^{1}$\BESIIIorcid{0000-0002-3935-619X},
M.~N.~Achasov$^{4,b}$\BESIIIorcid{0000-0002-9400-8622},
P.~Adlarson$^{81}$\BESIIIorcid{0000-0001-6280-3851},
X.~C.~Ai$^{86}$\BESIIIorcid{0000-0003-3856-2415},
R.~Aliberti$^{39}$\BESIIIorcid{0000-0003-3500-4012},
A.~Amoroso$^{80A,80C}$\BESIIIorcid{0000-0002-3095-8610},
Q.~An$^{77,64,\dagger}$,
Y.~Bai$^{62}$\BESIIIorcid{0000-0001-6593-5665},
O.~Bakina$^{40}$\BESIIIorcid{0009-0005-0719-7461},
Y.~Ban$^{50,g}$\BESIIIorcid{0000-0002-1912-0374}
% ...
\\
\vspace{0.2cm}
(BESIII Collaboration)\\
\vspace{0.2cm} {\it
$^{1}$ Institute of High Energy Physics, Beijing 100049, People's Republic of China\\
% ...
$^{4}$ Budker Institute of Nuclear Physics SB RAS (BINP), Novosibirsk 630090, Russia\\
% ...
$^{39}$ Johannes Gutenberg University of Mainz, Johann-Joachim-Becher-Weg 45, D-55099 Mainz, Germany\\
$^{40}$ Joint Institute for Nuclear Research, 141980 Dubna, Moscow region, Russia\\
% ...
$^{50}$ Peking University, Beijing 100871, People's Republic of China\\
% ...
$^{62}$ Southeast University, Nanjing 211100, People's Republic of China\\
% ...
$^{64}$ State Key Laboratory of Particle Detection and Electronics, Beijing 100049, Hefei 230026, People's Republic of China\\
% ...
$^{77}$ University of Science and Technology of China, Hefei 230026, People's Republic of China\\
% ...
$^{80}$ University of Turin and INFN, (A)University of Turin, I-10125, Turin, Italy; (B)University of Eastern Piedmont, I-15121, Alessandria, Italy; (C)INFN, I-10125, Turin, Italy\\
$^{81}$ Uppsala University, Box 516, SE-75120 Uppsala, Sweden\\
% ...
$^{86}$ Zhengzhou University, Zhengzhou 450001, People's Republic of China\\
\vspace{0.2cm}
$^{\dagger}$ Deceased\\
% ...
$^{b}$ Also at the Novosibirsk State University, Novosibirsk, 630090, Russia\\
% ...
$^{g}$ Also at State Key Laboratory of Nuclear Physics and Technology, Peking University, Beijing 100871, People's Republic of China\\
% ...
}
       }

\date{\today}

\begin{abstract}
Using a 10.9 fb$^{-1}$ data sample collected by the BESIII detector
 at center-of-mass energies from 4.16 to 4.34 GeV, we search for the charmless decays
 $X(3872) \to K_{S}^{0}K^{\pm}\pi^{\mp}$ and 
 $K^*(892)\bar{K}$, where the $X(3872)$ is produced via the radiative process $\ee \to \gamma X(3872)$. No significant signal is observed.
 We set upper limits on the relative branching fractions  
  $\mathcal{B}[\x\to\Ks\kpm \pi^{\mp}]/\mathcal{B}[\x\to\pip\pim\jpsi] <0.07$ and
$\mathcal{B}[\x\to K^*(892)\bar{K}]/\mathcal{B}[\x\to \pip\pim\jpsi] <0.10$  
 at the 90$\%$ confidence level.
 Additionally, upper limits on the product of the cross section $\sigma[\ee\to\gamma\x]$
 and the branching fractions $\mathcal{B}[\x\to K_{S}^{0}K^{\pm}\pi^{\mp}]$ and 
 $\mathcal{B}[\x\to K^*(892)\bar{K}]$ are reported at each energy point. In all cases, $K^*(892)\bar{K}$ refers to the sum of the modes $K^*(892)^+K^{-}+\text{c.c.}$ and $K^*(892)^0\bar{K}^0+\text{c.c.}$, where c.c. denotes the corresponding charge-conjugate modes. 

\end{abstract}

\newcommand{\BESIIIorcid}[1]{\href{https://orcid.org/#1}{\hspace*{0.1em}\raisebox{-0.45ex}{\includegraphics[width=1em]{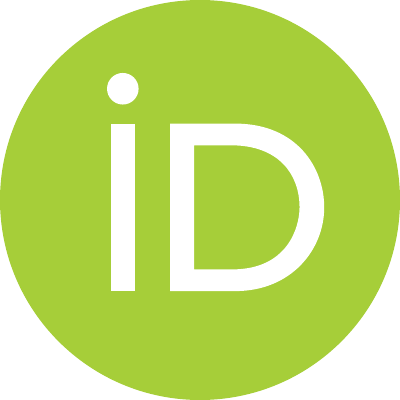}}}}

\maketitle

\section{INTRODUCTION}
The $X(3872)$, which is a strong candidate for an exotic hadron, was first observed by the Belle Collaboration
in 2003 in the decay $B^{\pm} \rightarrow K^{\pm}X(3872)$ with $X(3872) \to \pi^{+}\pi^{-}J/\psi$~\cite{belle_x3872},
 and was subsequently confirmed by CDF~\cite{cdf_x3872}, D0~\cite{d0_x3872},
 $BABAR$~\cite{babar_x3872}, LHCb~\cite{lhcb_x3872} and BESIII~\cite{ppjpsi}.
After more than 20 years of study, it is established as a narrow resonance with
 a Breit-Wigner width of $(1.19\pm0.21)$~MeV$/c^2$~\cite{2024pdg},  a mass of $(3871.64\pm0.06)$~MeV~(which is very close to the $D^{0}\bar{D}^{\ast 0}$ threshold), and quantum numbers
 $J^{PC}=1^{++}$~\cite{quantum_number}.  
 Several decay channels have been well measured, including   
  $\gamma J/\psi$, $\pip\pim J/\psi$, and $\omega J/\psi$, and it has also been found to strongly
 couple to $D^{0}\bar{D}^{\ast 0}$~\citep{DD_belle, DD_babar, Belle:2023zxm, BESIII:2020nbj}.
All of these observed decay modes of the $X(3872)$ contain a charm quark and an anticharm quark in the final state, suggesting it has a minimal quark content of $c\bar{c}$.
 Given its measured quantum numbers and mass, the closest match in the potential model is the $\chi_{c1}(2P)$ state. However, the mass of the $\x$ is lower than theoretical
 predictions for the $\chi_{c1}(2P)$, and significant isospin violation is observed in its
 decays, with the ratio
 $\mathcal{B}[\x\to\pip\pim\pi^0\jpsi]/\mathcal{B}[\x\to\pip\pim\jpsi]
 =1.6^{+0.4}_{-0.3} \pm 0.2$~\cite{isr_gammaX, lhcb_pipijpsi}.
 These features are unexpected for a pure $c\bar{c}$ state,  indicating that the $X(3872)$ may possess
 a more complex substructure, making it a candidate for an exotic hadron. Currently, 
various theoretical explanations exist for its nature, including
 interpretations as a charmonium state~\citep{Barnes:2003vb, Eichten:2004uh, Barnes:2005pb}, a hadronic molecule~\citep{molecule-like1, molecule-like2, molecule-like3}, a tetraquark~\citep{tetraquark1, tetraquark2, tetraquark3}, or a mixture of these 
 components~\citep{hybrid1, hybrid2, hybrid3}. 

  Investigating the decay patterns of the $\x$, such as the possible role of charmless decays, is helpful for
understanding its underlying structure.  
The decay with a final state of $K_{S}^{0}K^{\pm}\pi^{\mp}$ is a charmless light hadron decay and has been extensively studied in various charmonium decays  
~\citep{BaBar:2017dwm, LHCb:2023evz,BES:2006yed, BESIII:2024now,
  BESIII:2016dda},
  so an experimental measurement of $X(3872)\to K_{S}^{0}K^{\pm}\pi^{\mp}$ is 
  valuable.
  Additionally, the decays $X(3872)\to K^*(892)\bar{K}$ are charmless 
  $VP$ decays, where $V$ and $P$ denote vector and 
 pseudoscalar mesons, respectively. Assuming the $X(3872)$ is a $\bar{D}D^*+\text{c.c.} $ molecular bound state, which includes both neutral components~($\bar{D}^0D^{*0}+\text{c.c.}$) and charged components~($D^-D^{*+}+\text{c.c.}$), a theoretical prediction~\cite{kst_theo} based on 
intermediate $\bar{D}D^*+\text{c.c.}$ meson loops
estimates the branching fractions for $\x\to VP$ decay modes.
The numerical results are highly sensitive to the model parameter $\alpha$, which is dimensionless and calibrates the cutoff scale of the phenomenological form factors determining the transition amplitudes. The value of $\alpha$ is typically constrained by comparing the theoretical predictions with the corresponding experimental measurements.
In particular, the 
 $X(3872)\to K^*(892)\bar{K}$ branching fraction is calculated to be in a range of
$(0.08\text{--}3.99)\times 10^{-2}$~(with pure neutral components), 
 $(0.17\text{--}8.65)\times 10^{-2}$~(with the neutral components dominant) 
 or $(0.24\text{--}13.29)\times 10^{-2}$~(with equal neutral and charged components)~\cite{kst_theo}.

This analysis searches for the processes
 $\ee \to \gamma\x\to\gamma(\Ks\kpm\pi^{\mp})$ and
 $\ee \to \gamma\x\to\gamma(K^*(892)\bar{K})$, where $K^*(892)\bar{K}$ refers to the combinations
 $K^*(892)^+K^{-}+\text{c.c.}$ and $K^*(892)^0\bar{K}^0+\text{c.c.}$. 
  We analyze a data sample collected by the BESIII detector at 
center-of-mass~(c.m.) energies ranging
from 4.16 to 4.34 GeV with 
 a total integrated luminosity of 10.9 $\text{fb}^{-1}$~\cite{lum}.
The neutral state $K^*(892)^0(\bar{K}^*(892)^0)$ is reconstructed through its decay to 
$K^+\pim(K^-\pip)$, while the charged state 
$K^*(892)^+(K^*(892)^-)$ 
is reconstructed via $K^0\pip(\bar{K}^0\pim)$.
Both decay channels, namely 
$X(3872)\to K_{S}^{0}K^{\pm}\pi^{\mp}$ and $X(3872)\to K^*(892)\bar{K}$, 
yield the same final state $K_{S}^{0}K^{\pm}\pi^{\mp}$. 
The $K_{S}^{0}$ is reconstructed via its decay to $\pip \pim$.

\section{BESIII DETECTOR AND MONTE CARLO SIMULATION}
The BESIII detector~\cite{Ablikim:2009aa} records symmetric $e^+e^-$ collisions 
provided by the BEPCII storage ring~\cite{Yu:IPAC2016-TUYA01}
in the c.m. energy range from 1.84 to 4.95~GeV, 
with a peak luminosity of $1.1 \times 10^{33}\;\text{cm}^{-2}\text{s}^{-1}$ 
achieved at c.m.\ energy $\sqrt{s} = 3.773\;\text{GeV}$. 
BESIII has collected large data samples in this energy region~\cite{Ablikim:2019hff}.
 The cylindrical core of the BESIII detector covers 93\% of the full solid angle and consists of a helium-based
 multilayer drift chamber~(MDC), a time-of-flight
system~(TOF), and a CsI(Tl) electromagnetic calorimeter~(EMC),
which are all enclosed in a superconducting solenoidal magnet
providing a 1.0~T magnetic field.
The solenoid is supported by an
octagonal flux-return yoke with resistive plate counter muon
identification modules interleaved with steel. 
The charged-particle momentum resolution at $1~{\rm GeV}/c$ is
$0.5\%$, and the 
${\rm d}E/{\rm d}x$
resolution is $6\%$ for electrons
from Bhabha scattering. The EMC measures photon energies with a
resolution of $2.5\%$ ($5\%$) at $1$~GeV in the barrel (end cap)
region. The time resolution in the plastic scintillator TOF barrel region is 68~ps, while
that in the end cap region was 110~ps. The end cap TOF
system was upgraded in 2015 using multigap resistive plate chamber
technology, providing a time resolution of
60~ps, which benefits 83\% of the data used in this analysis~\cite{etof}.

The Monte Carlo (MC) simulation samples are generated using a GEANT4-based 
software package~\cite{geant4}, incorporating a detailed geometric description
 of the BESIII detector and its response. These simulations are crucial for 
evaluating detection efficiencies, optimizing event selection criteria, and
 estimating background contributions.
 At each energy point, 100,000 signal events are generated
 for both processes $\ee\to\gamma\x\to\gamma\Ks\kpm\pimp$ and
  $\ee\to\gamma\x\to\gamma K^*(892)\bar{K}$. 
The $\ee\to\gamma\x$
is modeled as an E1 radiative transition, consistent with
 BESIII experimental data~\cite{ppjpsi}. 
 Initial state radiation (ISR) is simulated with {\sc kkmc}~\cite{ref:kkmc}, by 
 incorporating the $\sqrt{s}$-dependent 
 production cross section of $\ee\to\gamma\x$ 
 into the program~\cite{isr_gammaX}, where 
the energies of ISR photons range up to the values at the production threshold of the $\gamma\x$ system. 
 
 The $\x\to\Ks K^{\pm}\pi^{\mp}$ and $\x\to K^*(892)\bar{K}$ decays are 
 described with the phase-space model in {\sc evtgen}~\cite{ref:evtgen}. Subsequent decays, $K^*(892)^0(\bar{K}^*(892)^0)\to\kp\pim(\km\pip)$ and 
 $K^*(892)^+(K^*(892)^-)\to K^0\pip(\bar{K}^0\pim)$, are modeled with the VSS model in 
 {\sc evtgen} which describes the decay of a vector particle into two scalars.  
 The $K^*(892)^+K^-+\text{c.c.}$ and $K^*(892)^0\bar{K}^0+\text{c.c.}$ processes are 
 simulated together and are assumed to be produced at the same rate, which is consistent with isospin conservation. 
 The decay $\Ks\to\pip\pim$ is also modeled using 
 the phase-space approach.
 Final state radiation is simulated with {\sc photos}~\cite{photos2}.

 To investigate potential background contributions, inclusive MC samples are 
 utilized. These samples encompass open charm processes, ISR production of vector 
 charmonium(like) states, and continuum processes, all simulated with
 {\sc kkmc}~\cite{ref:kkmc}. Particle decays are generated using {\sc evtgen}~\cite{ref:evtgen},
 with branching fractions sourced from the Particle Data Group~\cite{2024pdg} when 
 available, and otherwise modeled 
 with {\sc lundcharm}~\cite{ref:lundcharm} for charmonium states 
 and {\sc pythia}~\cite{pythia} for other hadrons.

\section{EVENT SELECTION AND BACKGROUND ANALYSIS}

 Charged tracks detected in the MDC are required to be within a polar angle ($\theta$)
  range of $|\rm{cos\theta}|<0.93$, where $\theta$ is defined with respect to the
  $z$ axis, which is the symmetry axis of the MDC. Particle identification~(PID) for charged tracks combines measurements of the energy
 deposited in the MDC~(d$E$/d$x$) and the flight time in the TOF to form likelihoods
 $\mathcal{L}(h)$ for each hadron hypothesis $h=p, K,\pi$.
 A track is classified as a kaon if $\mathcal{L}(K)>\mathcal{L}(p)$ and 
 $\mathcal{L}(K)>\mathcal{L}(\pi)$.
Identified kaons must satisfy vertex requirements:
 $V_r < 1.0$ cm and $|V_{z}| < 10.0$ cm, 
 where $V_{r}$ and $V_{z}$ are the closest
 distances of charged tracks to the interaction point (IP) in the plane perpendicular
 to and along the  beam direction, respectively.
 All remaining charged tracks are treated as pion candidates. Selected events must 
 contain exactly one identified kaon and three pion candidates, with a total charge
 of the four tracks summing to zero.

Photon candidates are identified from isolated showers in
 the EMC. Each shower must have a deposited energy exceeding 25 MeV in
 the barrel region ($|\cos \theta|< 0.80$) or 50 MeV in
 the end cap region ($0.86 <|\cos \theta|< 0.92$). To exclude showers produced 
 by charged tracks, the angle between the EMC
 shower and the extrapolated position of the closest charged
 track at the EMC, measured from the IP,
 must be greater than 10 degrees. To suppress electronic noise
 and showers unrelated to the collision event, the difference between the 
 EMC shower time and the event start time must lie
 within [0, 700] ns. Events with at least one photon candidate
 meeting the above criteria are selected.

The $K_{S}^0$ candidates are reconstructed from pairs of oppositely
 charged pion tracks ($\pi^{+}\pi^{-}$) selected from the three pion
 candidates. 
 All such pairs are subjected to a secondary vertex fit, ensuring 
 the tracks originate 
 from a common vertex. The invariant mass of these pairs must satisfy  
 $|M_{\pi^{+}\pi^{-}} - m_{K_{S}^{0}}|<$ 22~MeV$/c^{2}$, where
$M_{\pi^{+}\pi^{-}}$ denotes the invariant mass of the $\pi^{+}\pi^{-}$ combination and 
$m_{K_{S}^{0}}$ is the $K^0_{S}$ nominal mass~\cite{2024pdg}.
 The decay length of the $K_{S}^0$ candidate, measured relative to the IP, 
 must exceed twice the resolution of the vertex position.
 If more than one $\pi^{+}\pi^{-}$ combination in an event satisfies the above criteria, the combination with the least 
  secondary vertex fit chi-square value is retained for further analysis.
To estimate the background from
 non-$K_{S}^0$ processes, we use the events in the $K_{S}^0$ sideband regions defined by
$0.432 < M_{\pip\pim} < 0.454$~GeV$/c^{2}$ and
 $0.542 < M_{\pip\pim} < 0.564$~GeV$/c^{2}$.
 After reconstructing the $K_{S}^0$, one pion candidate remains, 
 which must satisfy the vertex constraints
  $V_r < 1.0$ cm and $|V_{z}| < 10.0$ cm.

Signal event candidates are identified as those containing four charged 
tracks, specifically one kaon, one additional pion, and the two pions originating from the $K_{S}^0$, as well as at least one photon candidate. 
A vertex fit is further performed on the kaon, the additional pion, and the $K_{S}^0$ to 
confirm they originate from a common vertex at the IP. 
To enhance momentum resolution and suppress backgrounds, a 
four-momentum constraining kinematic
 fit (4C-fit) is applied to each signal event candidate, enforcing energy and 
momentum conservation with the initial $\ee$ collision system. 
Only events with a $\chi_{\text{4C}}^{2}$ value less than 50 are retained, where $\chi^2_{\text{4C}}$ represents the chi-square value from the 4C-fit for the signal hypothesis.
In events with multiple possible combinations (e.g., multiple photons), the combination yielding the smallest $\chi_{\text{4C}}^{2}$ is
 selected for further analysis.

 To suppress backgrounds from processes without a photon, such as 
 $e^+e^-\rightarrow K_S^0K\pi$, a 4C fit is performed to the background hypothesis with four tracks and no photon, and the resulting  $\chi^2_{\text{4C},-\gamma}$ is required to be greater than $\chi^2_{\text{4C}}$.
Similarly, to suppress backgrounds with an additional photon, we 
 impose $\chi^2_{\text{4C}}<\chi^2_{\text{4C},+\gamma}$, where $\chi^2_{\text{4C},+\gamma}$ denotes the 
 chi square from a fit for a hypothesis with an extra photon.

Background events containing a $\pi^0$ arise from processes such as 
$\ee\to\gamma^{\text{ISR}}\pi^0\Ks K^{\pm}\pi^{\mp}$ and $\ee\to\gamma^{\text{ISR}}\rho^+K^0K^-+\text{c.c.}$,  
with $\rho^+\to\pi^0\pip$ and $K^0\to\Ks$. 
For the signal process $\x\to\Ks K^{\pm}\pi^{\mp}$, a photon recoils against the $\Ks K^{\pm}\pi^{\mp}$ system, and thus the invariant mass squared $M_{\text{recoil}}^{2}(K_{S}^{0}K^{\pm}\pi^{\mp})$ of the recoil system is expected to be zero.
To eliminate the remaining $\pi^0$ backgrounds, we require that 
$M_{\text{recoil}}^{2}(K_{S}^{0}K^{\pm}\pi^{\mp}) < 0.012$~(GeV/$c^2$)$^{2}$.
 
  The decay channels $X(3872)\rightarrow\Ks K^{\pm}\pi^{\mp}$ and 
 $X(3872)\rightarrow K^*(892)\bar{K}$ share the same final state of $\Ks K^{\pm}\pi^{\mp}$, 
 and thus the same event selection criteria described above applies to both channels.
 For the $X(3872)\rightarrow K^*(892)\bar{K}$ decay, additional requirements
 are imposed to select $K^{*}(892)$ candidates.
 Specifically, the invariant mass of the $K^{\pm}\pi^{\mp}$ pair must satisfy
 $0.748 < M_{K^{\pm}\pi^{\mp}} < 1.036$~GeV$/c^{2}$~[three times the $K^*(892)^0$ width] to identify $K^*(892)^0$ or $\bar{K}^*(892)^0$ candidates, or the invariant mass of the $\Ks\pi^{\pm}$ pair must fall within $0.743 < M_{\Ks\pi^{\pm}} <1.05$~GeV$/c^{2}$~[three times the $K^*(892)^{\pm}$ width] to select 
 $K^*(892)^{\pm}$  candidates. Figure~\ref{kskpi} displays the invariant mass distribution of 
 the $\Ks K^{\pm}\pi^{\mp}$ and $K^*(892)\bar{K}$ systems for all datasets combined after
 applying the event selection criteria.

\section{SIGNAL EXTRACTION AND CROSS SECTION MEASUREMENT}
\begin{figure}
	\centering
	\includegraphics[width=0.4\textwidth]{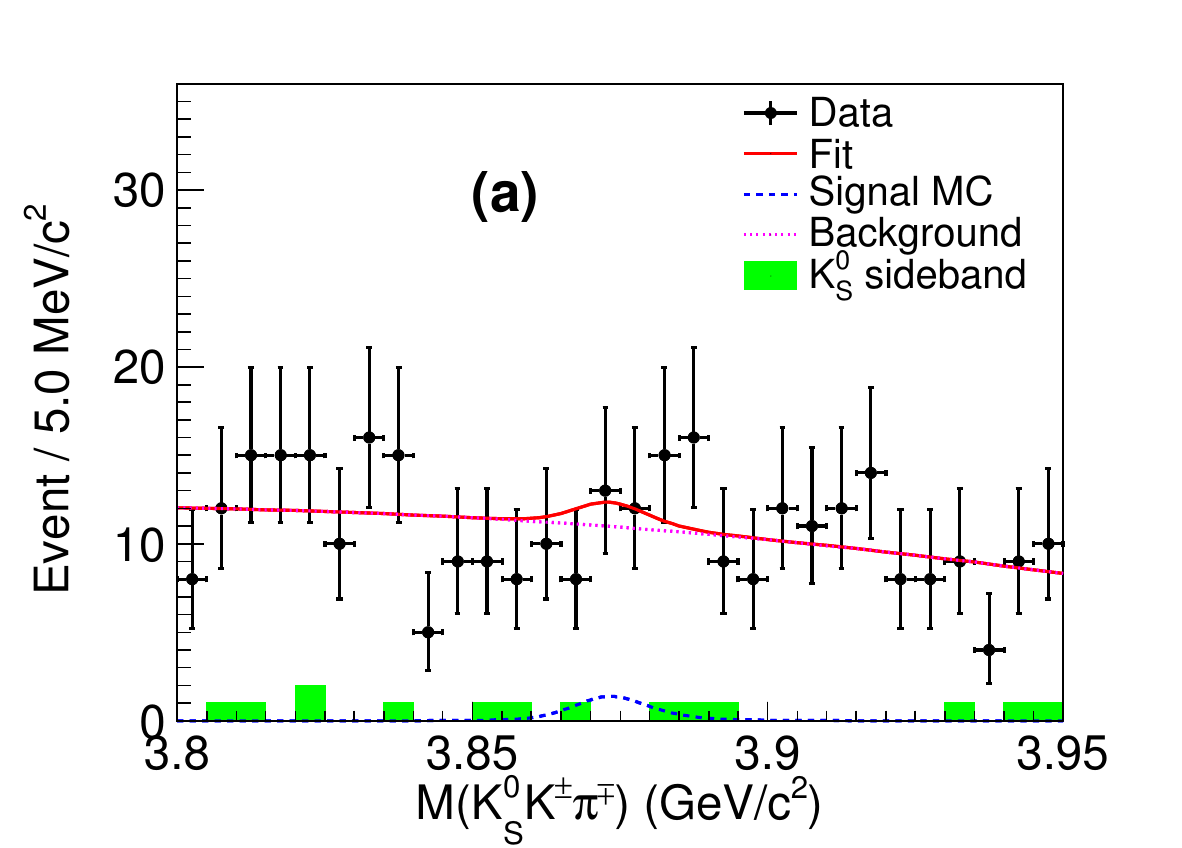}
	\includegraphics[width=0.4\textwidth]{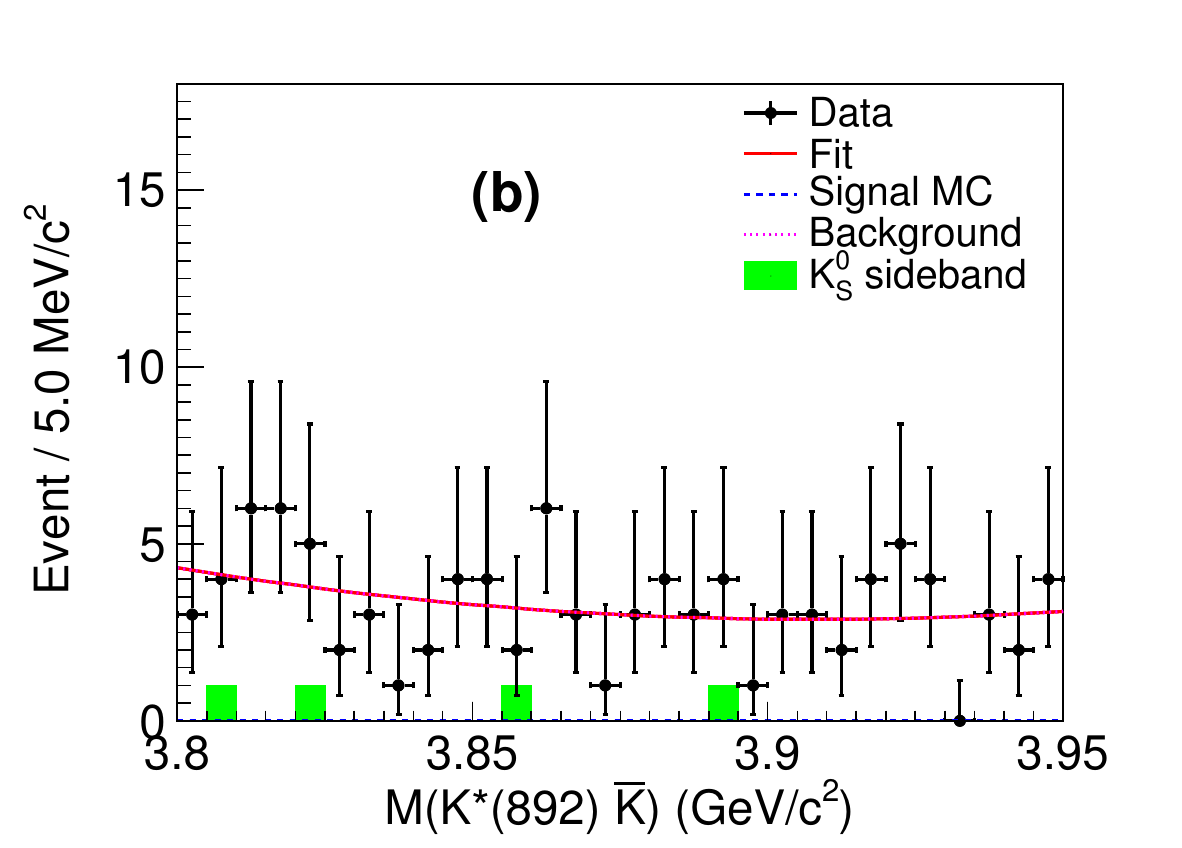}
	\vspace{-5mm}
	\caption{Fits to the invariant mass distributions of $M(\Ks K^{\pm}\pi^{\mp})$ and $M(K^*\bar{K})$ from data for the channels (a) $\x\to\Ks K^{\pm}\pi^{\mp}$ and (b) $\x\to K^*(892)\bar{K}$. 
The dots with error bars represent the full dataset for all c.m. energies combined. The blue dashed curves are the signal PDFs, the red solid curves are the total fit results,
    the pink dashed curves represent the background components, and the green shaded histograms show the background estimated from the $\Ks$ sideband region~(in the lower plot, the fitted signal yield is negligible, rendering the signal PDFs essentially invisible and causing the background to nearly coincide with the total fit; consequently, the blue and pink curves are barely distinguishable).}
	\label{kskpi}
\end{figure}

To extract the signal yield, an extended unbinned maximum likelihood fit is applied to the
 invariant mass distribution of $\Ks K^{\pm}\pi^{\mp}$ or $K^*(892)\bar{K}$.
The signal probability density function~(PDF) is modeled by convolving the $X(3872)$ signal shape, 
derived from the MC simulation, with a Gaussian function. This Gaussian function 
accounts for the 
mass resolution differences between the MC simulation and the data. 
Its mean and width parameters are fixed to values determined from a control sample of 
$\psi(2S)\to\gamma\chi_{c1}$, with $\chi_{c1}\to\Ks K^{\pm}\pi^{\mp}$, sharing the same 
final state as the signal process. The background is described using a second-order 
Chebychev polynoimal function. Figure~\ref{kskpi} displays the fitted curves overlaid on the data for the decay channels $\x\to\Ks K^{\pm}\pi^{\mp}$ and $\x\to K^*(892)\bar{K}$, with the fit results showing no significant signal for the $X(3872)$ in either channel.


The upper limit at the 90$\%$ confidence level~(CL) for the signal yield of 
$X(3872)$ at each c.m.\ energy, $N^{\text{UL}}$, is calculated 
employing a Bayesian approach~\cite{bayesian1}.
 This method involves scanning the likelihood function over assumed signal
 event counts from zero to a sufficiently large upper bound, and defining the upper
 limit as the point where the integrated likelihood reaches 90$\%$ of the total 
 probability. By excluding negative signal yields,
 this approach ensures a conservative estimation. These upper limits on the signal 
 yields are subsequently converted into corresponding upper limits on the product
 of the cross section $\sigma[\ee\to\gamma\x]$
  and branching fractions $\mathcal{B}[\x\to K_{S}^{0}K^{\pm}\pi^{\mp}]$ or
  $\mathcal{B}[\x\to K^*(892)\bar{K}]$ using the following expression:
\begin{linenomath*}  
\begin{align}
\sigma^{\text{UL}}[\ee\to\gamma\x] \cdot \mathcal{B}[&\x\to K_{S}^{0}K^{\pm}\pi^{\mp}/K^*\bar{K}] \notag \\
&= \frac{N^{\text{UL}}}
{\mathcal{L}_{\rm int}(1+\delta)\varepsilon\mathcal{B}}.
\end{align}
\end{linenomath*}
 Here, $\varepsilon$ is the selection efficiency; 
for decay channel $\x\to K_{S}^{0}K^{\pm}\pi^{\mp}$, $\mathcal{B}$ is
 the branching fraction for $\Ks\to\pip\pim$;  for decay channel $\x\to K^*(892)\bar{K}$, 
 $\mathcal{B}$ is a product of branching fractions
 involving the decays in the $K^*(892)\bar{K}$ channel: 
 $K^*(892)^+(K^*(892)^-)\to K^0\pip(\bar{K}^0\pim)$,
  $K^*(892)^0(\bar{K}^*(892)^0)\to K^+\pim(K^-\pip)$, 
$K^0(\bar{K}^0)\to\Ks$, and $\Ks\to\pip\pim$; 
$\mathcal{L}_{\rm int}$ is the integrated luminosity; and $(1+\delta)$ accounts for the radiative corrections~\cite{isr_gammaX, Sun:2020ehv}.

Table~\ref{count_UL} provides 
 the upper limits on the product of
 the cross section and branching fractions for the decay channels
  $\x\to K_{S}^{0}K^{\pm}\pi^{\mp}$ and $\x\to K^*(892)\bar{K}$, at each c.m. energy.  
  Systematic uncertainties have been incorporated
 into the upper limit calculations.

\begin{table*}[ht]
	\caption{Upper limits at 90$\%$ CL for the measurement of the product of the 
  cross section and 
  branching ratio, $\sigma[\ee\to\gamma\x]\cdot\mathcal{B}[\x\to\Ks\kpm\pi^{\mp}]$
  and $\sigma[\ee\to\gamma\x]\cdot\mathcal{B}[\x\to K^*(892)\bar{K}]$, at different c.m. 
   energies. The table includes the integrated luminosity
  $\mathcal{L}_{\text{int}}$, upper limits on the number of signal events 
  $N^{\text{UL}}_1$ and $N^{\text{UL}}_2$~[where the subscript 1 denotes the $\x\to\Ks\kpm\pimp$ process, and 
   the subscript 2 represents the $\x\to K^*(892)\bar{K}$ process], detection 
  efficiencies $\varepsilon_1$ and $\varepsilon_2$, 
  radiative correction factor $(1+\delta)$, upper limits on 
  $\sigma_1\cdot\mathcal{B}_1$ and 
  $\sigma_2\cdot\mathcal{B}_2$. The last column indicates the data-taking time corresponding to each data sample. 
  Systematic uncertainties are incorporated into the upper limits.}

	\hspace{15pt}
	\centering
	\begin{tabular}{c c c c c c c c c c} 
		\hline 
		\hline
		 $E_{\text{c.m.}}(\text{GeV})$ & $\mathcal{L}_{\text{int}}$($\text{pb}^{-1}$) & $N^{\text{UL}}_1$ & $N^{\text{UL}}_2$ 
        & $\varepsilon_1(\%)$ & $\varepsilon_2(\%)$ & $(1+\delta)$ 
        & $\sigma_1\cdot\mathcal{B}_1$({\rm pb}) & $\sigma_2\cdot\mathcal{B}_2$({\rm pb}) & Year\\
		\hline
 4.157 & 408.2   &  < 5.8  & < 2.4   & 20.8 & 18.0 & 0.76 & < 0.13 & < 0.19 & 2019\\
 4.178 & 3194.5  & < 10.7 & < 4.6   & 21.8 & 20.8  & 0.76 & < 0.03 & < 0.04 & 2016\\
 4.189 & 526.7   & < 7.5  & < 4.0   & 21.9 & 20.8  & 0.77 & < 0.13 & < 0.21 & 2017\\
 4.199 & 526.0   & < 5.5  & < 3.9   & 21.5 & 20.8  & 0.78 & < 0.09 & < 0.20 & 2017\\
 4.209 & 517.1   & < 5.2  & < 3.8   & 21.0 & 19.9  & 0.80 & < 0.09 & < 0.20 & 2017\\
 4.219 & 514.6   & < 6.4  & < 5.4   & 20.7 & 19.5  & 0.82 & < 0.11 & < 0.29 & 2017\\
 4.226 & 1056.4  & < 5.3  & < 6.0   & 21.1 & 20.0  & 0.84 & < 0.04 & < 0.15 & 2013\\
 4.236 & 530.3   & < 5.1  & < 5.3   & 20.1 & 19.2  & 0.86 & < 0.08 & < 0.26 & 2017\\
 4.244 & 538.1   & < 3.1  & < 3.8   & 19.6 & 18.6  & 0.89 & < 0.05 & < 0.19 & 2017\\
 4.258 & 828.4   & < 5.2  & < 5.0   & 18.7 & 17.8  & 0.93 & < 0.06 & < 0.16 & 2013\\
 4.267 & 531.1   & < 7.4  & < 2.7   & 18.1 & 17.3  & 0.96 & < 0.12 & < 0.14 & 2017\\
 4.278 & 175.7   & < 4.0  & < 3.9   & 16.9 & 16.1  & 0.99 & < 0.20 & < 0.61 & 2017\\
 4.288 & 502.4   & < 5.1  & < 3.1   & 15.5 & 13.7  & 1.02 & < 0.10 & < 0.20 & 2019\\
 4.312 & 501.1   & < 2.7  & < 2.7   & 14.5 & 12.5  & 1.09 & < 0.05 & < 0.18 & 2019\\
 4.337 & 504.9   & < 7.1  & < 5.3   & 13.4 & 11.9  & 1.16 & < 0.13 & < 0.33 & 2019\\
		\hline 
		\hline
	\end{tabular}
	\label{count_UL}  
\end{table*}

\section{RELATIVE BRANCHING RATIO MEASUREMENT}
Since the decay channel $\x\to \pi^{+}\pi^{-} J/\psi$ has been well measured by the BESIII experiment~\cite{ppjpsi}, we also determine 
the relative branching ratios 
$\mathcal{R}_1\equiv\frac{\mathcal{B}[\x\to\Ks\kpm\pi^{\mp}]}
{\mathcal{B}[\x\to\pip\pim J/\psi]}$, 
and
$\mathcal{R}_2\equiv\frac{\mathcal{B}[\x\to K^*(892)\bar{K}]}{\mathcal{B}[\x\to\pip\pim\jpsi]}$
, which are calculated using the equations:
\begin{linenomath*}
\begin{equation}
\label{eq_relative_BF}
	\begin{aligned}
		\mathcal{R}_1 
        &=\frac{N_{\Ks\kpm\pimp}}{N_{\pip\pim\jpsi}} \cdot 
        \frac{\varepsilon^{\text{ave}}_{\pip\pim\jpsi}}{\varepsilon^{\text{ave}}_{\Ks\kpm\pimp}}
          \cdot \frac{\mathcal{B}(J/\psi \rightarrow l^{+}l^{-})}{\mathcal{B}} \\
	\end{aligned}
\end{equation}
\end{linenomath*}
and
\begin{linenomath*}
\begin{equation}
\label{eq_relative_BF}
	\begin{aligned}
		\mathcal{R}_2
        &=\frac{N_{K^*(892)\bar{K}}}{N_{\pip\pim\jpsi}} \cdot 
        \frac{\varepsilon^{\text{ave}}_{\pip\pim\jpsi}}{\varepsilon^{\text{ave}}_{K^*K}}
          \cdot \frac{\mathcal{B}(J/\psi \rightarrow l^{+}l^{-})}{\mathcal{B}}. \\
	\end{aligned}
\end{equation}
\end{linenomath*}
Here, $N_{\Ks\kpm\pimp}$ and $N_{K^*(892)\bar{K}}$
represent the total number of signal events for
 the decay channels $\x\to\Ks K^{\pm}\pi^{\mp}$ and $\x\to K^*(892)\bar{K}$, derived by fitting the combined data samples across all energies; 
 $N_{\pip\pim\jpsi}=86.3^{+10.5}_{-9.8}$ is the number of signal events in the 
 normalization channel $\x\to \pi^{+}\pi^{-} J/\psi$, following the approach of Ref.~\cite{isr_gammaX};
 $\varepsilon^{\text{ave}}_{\Ks\kpm\pimp}$ 
 and $\varepsilon^{\text{ave}}_{K^*(892)\bar{K}}$ denote the averaged detection efficiency for the decay channels $\x\to\Ks K^{\pm}\pi^{\mp}$ and $\x\to K^*(892)\bar{K}$; 
 $\varepsilon^{\text{ave}}_{\pip\pim\jpsi}$ is the averaged detection efficiency
  for the normalization channel $\x\to \pi^{+}\pi^{-} J/\psi$,  
calculated using efficiencies from Ref.~\cite{BESIII:2023eeb}; 
$\mathcal{B}(J/\psi \to l^{+}l^{-})$ is the branching fraction of $\jpsi\to l^{+}l^{-}$~( $l=e$ or $\mu$);
$\mathcal{B}$ represents the product of branching fractions for the specific decay 
channels, consistent with the cross section measurement. 
The averaged efficiencies ($\varepsilon^{\text{ave}}_{\Ks\kpm\pimp}$, $\varepsilon^{\text{ave}}_{K^*(892)\bar{K}}$ 
and $\varepsilon^{\text{ave}}_{\pip\pim\jpsi}$)  are calculated by weighting the 
efficiencies at each energy by the integrated luminosity and the production 
cross section of $\ee\to\gamma\x$.

 In the absence of a significant signal in the decay channels 
$\x\to\Ks K^{\pm}\pi^{\mp}$ and $\x\to K^*(892)\bar{K}$,  upper limits on the signal yields 
for the combined data samples are established at the 90$\%$ CL:
$N<23.0$ for $\x\to\Ks K^{\pm}\pi^{\mp}$ and 
 $N<10.7$ for $\x\to K^*(892)\bar{K}$.
Applying Eq.~\eqref{eq_relative_BF}, the corresponding upper limits on the
relative branching ratios are $\mathcal{R}_1 < 0.07$ and $\mathcal{R}_2 < 0.10$, 
respectively, with systematic uncertainties considered.

\section{SYSTEMATIC UNCERTAINTY}   
  The systematic uncertainties in the measurement of the cross section arise from 
 multiple sources, including the integrated luminosity, photon detection efficiency, 
 $\Ks$ reconstruction, charged track reconstruction, PID, kinematic fit, MC decay model, 
 radiative correction, background veto, branching fractions, and signal extraction. 
The systematic uncertainties mentioned above can be broadly classified into two 
 categories: additive and multiplicative. The signal extraction is classified as an additive systematic uncertainty, while the remaining sources are categorized as 
 multiplicative  systematic uncertainties.

The integrated luminosity is determined using large-angle Bhabha scattering events, 
with a systematic uncertainty of 0.7$\%$~\cite{lum}.

The photon detection efficiency contributes a systematic uncertainty  of 1.0$\%$, 
as established in Ref.~\cite{photon_sys}.

The systematic uncertainty associated with the $\Ks$ reconstruction efficiency 
is assessed using control samples 
$\jpsi\to K^{*\pm}K^{\mp}$ and $\jpsi\to\Ks K^{\pm}\pi^{\mp}$~\cite{ks_sys}.
This uncertainty, which includes the pion tracking efficiency  and the $\Ks$ mass window selection, is quantified as 1.2$\%$ based on differences between data and MC simulation.

The tracking efficiency for charged particles originating from the IP 
introduces a systematic uncertainty. For the single kaon and pion tracks, 
each contributes a 1.0$\%$ uncertainty, yielding a total tracking uncertainty of 
2.0$\%$~\cite{pi_k_sys_report}. 

The PID is applied solely to the kaon track, with a systematic uncertainty of 
1.0$\%$ in the PID efficiency, as reported in Ref.~\cite{K_PID_sys}. 
   
\begin{table*}[ht]
	\caption{ Systematic uncertainties (in percentage) for the measurement of 
  $\sigma(\ee\to\gamma\x)\cdot\mathcal{B}(\x\to\Ks\kpm$  $\pi^{\mp})$ and
  $\sigma(\ee\to\gamma\x)\cdot\mathcal{B}(\x\to K^*(892)\bar{K})$ (values in parentheses).
  The total systematic uncertainty is the quadrature sum of individual contributions.}
	\hspace{15pt}
	\centering
	\begin{tabular}{c c c c c c c c c c c c} 
		\hline 
		\hline
		$E_{\text{c.m.}}$ (GeV) & Luminosity & Photon & Tracking  & PID  & Kinematic fit
        & $(1+\delta)$ & MC model  
        & $K^{0}_{S}$ rec. & $\pi^0$ veto & Total~($\%$)\\
		\hline
4.157 & 0.7  & 1.0   & 2.0  & 1.0 & 1.7 (1.2) & 0.1 & 0.4 (5.3) & 1.2 & 4.3 (3.2) & 5.4 (6.9)\\
4.178 & 0.7  & 1.0   & 2.0  & 1.0 & 0.7 (1.0) & 0.8 & 2.4 (4.6) & 1.2 & 4.3 (3.2) & 5.2 (6.3)\\
4.189 & 0.7  & 1.0   & 2.0  & 1.0 & 0.9 (0.9) & 1.1 & 0.4 (5.0) & 1.2 & 4.3 (3.2) & 4.8 (6.3)\\
4.199 & 0.7  & 1.0   & 2.0  & 1.0 & 1.2 (1.1) & 0.9 & 2.5 (5.8) & 1.2 & 4.3 (3.2) & 5.6 (7.0)\\
4.209 & 0.7  & 1.0   & 2.0  & 1.0 & 1.2 (1.2) & 0.9 & 1.5 (4.9) & 1.2 & 4.3 (3.2) & 5.3 (6.2)\\
4.219 & 0.7  & 1.0   & 2.0  & 1.0 & 1.0 (1.2) & 1.0 & 0.3 (5.0) & 1.2 & 4.3 (3.2) & 4.9 (6.3)\\
4.226 & 0.7  & 1.0   & 2.0  & 1.0 & 1.1 (1.2) & 0.8 & 1.1 (5.7) & 1.2 & 4.3 (3.2) & 5.0 (6.6)\\
4.236 & 0.7  & 1.0   & 2.0  & 1.0 & 1.0 (1.1) & 1.0 & 1.2 (4.7) & 1.2 & 4.3 (3.2) & 5.1 (6.0)\\
4.244 & 0.7  & 1.0   & 2.0  & 1.0 & 1.2 (1.1) & 0.9 & 1.8 (4.2) & 1.2 & 4.3 (3.2) & 5.4 (5.7)\\
4.258 & 0.7  & 1.0   & 2.0  & 1.0 & 1.2 (1.1) & 0.9 & 2.0 (6.0) & 1.2 & 4.3 (3.2) & 5.5 (6.9)\\
4.267 & 0.7  & 1.0   & 2.0  & 1.0 & 1.3 (1.1) & 1.1 & 0.9 (4.8) & 1.2 & 4.3 (3.2) & 5.0 (6.0)\\
4.278 & 0.7  & 1.0   & 2.0  & 1.0 & 1.0 (1.3) & 1.0 & 2.9 (2.8) & 1.2 & 4.3 (3.2) & 5.7 (5.0)\\
4.288 & 0.7  & 1.0   & 2.0  & 1.0 & 1.5 (1.7) & 0.9 & 0.4 (5.0) & 1.2 & 4.3 (3.2) & 5.4 (6.6)\\
4.312 & 0.7  & 1.0   & 2.0  & 1.0 & 1.7 (1.2) & 0.6 & 0.8 (3.6) & 1.2 & 4.3 (3.2) & 5.5 (5.3)\\
4.337 & 0.7  & 1.0   & 2.0  & 1.0 & 1.2 (1.4) & 0.6 & 1.2 (4.0) & 1.2 & 4.3 (3.2) & 5.4 (5.6)\\
		\hline 
		\hline
	\end{tabular}
	\label{sys}  
\end{table*}

To address discrepancies in track helix parameters between data and MC simulations, 
a correction method is implemented during the kinematic fit, following 
Ref.~\cite{helix_correction}. The systematic uncertainty from the kinematic fit is 
conservatively estimated as the difference in detection efficiencies between 
corrected and uncorrected MC samples.

For the $\x\to\Ks K^{\pm}\pi^{\mp}$ decay channel, the nominal MC model employs a 
phase space distribution. An alternative model, inspired by the decay of the
$\chi_{c1}$ state~\cite{2024pdg} with identical $J^{PC}=1^{++}$ quantum numbers, assumes the $X(3872)$
 decays via intermediate $K^*(892)\bar{K}$ resonance states, subsequently
 producing the $\Ks K^{\pm}\pi^{\mp}$ final state. 
 The systematic uncertainty due to the
 MC decay model is determined by comparing detection efficiencies between the 
 nominal phase space model and this alternative resonance-based model.

 Similarly, for the $\x\to K^*(892)\bar{K}$ channel, which also results in the 
 $\Ks K^{\pm}\pi^{\mp}$ final state, the nominal model assumes an equal ratio of $K^*(892)^+K^- +\text{c.c.}$ and $K^*(892)^0\bar{K}^0+\text{c.c.}$, as predicted by isospin symmetry. 
 The alternative resonance-based approach adopts the Body3 model~\cite{ref:evtgen}, incorporating the possible 
 decay dynamics of the $\chi_{c1}$ state and accounting for intermediate $K^*(892)$ 
 resonances. This model generates 
 three-body $\Ks\kpm\pimp$ events using the Dalitz plot and
 angular distributions from $\chi_{c1}$ data. The systematic uncertainty is evaluated 
 as the difference in detection efficiencies between the nominal and alternative models.

The radiative correction factor is computed using the method in Ref.~\cite{Sun:2020ehv}, 
derived from the cross section line shape of $\ee \to\gamma\x$~\cite{isr_gammaX}.
The line shape is modeled with a Breit-Wigner function for the $Y(4200)$ resonance, 
characterized by a mass of $M=4200.6^{+7.9}_{-13.3}\pm3.0~\text{MeV}/c^2$ 
 and a width of $\Gamma=115^{+38}_{-26}\pm12$ MeV.
To estimate the uncertainty in the radiative correction due to these 
resonance parameters, a two-dimensional Gaussian sampling method is employed.
 Specifically, 300 sets of mass and width parameters for the $Y(4200)$ are generated, 
 and for each set, the product $(1+\delta)\varepsilon$ is recalculated using the 
 weight method from Ref.~\cite{Sun:2020ehv}. The relative standard deviation of 
 the resulting  $(1+\delta)\varepsilon$ distribution,  expressed as a percentage, 
 is taken as the systematic uncertainty.

 The systematic uncertainty from the $\pi^0$ veto criterion, defined as
$M_{\text{recoil}}^{2}(K_{S}^{0}K^{\pm}\pi^{\mp}) < 0.012$~(GeV/$c^2$)$^{2}$,   
is related to the photon energy~(the efficiency loss induced by $\pi^0$ veto criterion exhibits a direct positive correlation with photon energy).
To estimate the systematic uncertainty associated with this criterion, 
control samples are selected from the $\psi(2S)\to\gamma\chi_{c1}$ 
 and $\ee\to\gamma^{\text{ISR}}\jpsi$~(at 
 3.773~GeV,  4.178~GeV and 4.226~GeV) processes. 
These processes provide photons with varying energies.
Efficiency differences between MC simulations and datasets
 are analyzed, and we extrapolate the systematic uncertainty to our studied signal process.

The background veto criterion  $\chi^2_{\text{4C}}<\chi^2_{\text{4C},\pm\gamma}$  introduces
 a systematic uncertainty, assessed using the control sample
$\psi(2S)\to\gamma\chi_{c1}\to\gamma\Ks\kpm\pi^{\mp}$. The efficiency difference 
between signal MC simulations and data is 0.1$\%$, and is deemed negligible.
Given the high precision of the branching fractions for $\Ks \to \pi^{+}\pi^{-}$
and relevant $K^*(892)$ decays~\cite{2024pdg},  their uncertainties are negligible 
and omitted from the total systematic uncertainty.

 Assuming the above systematic uncertainties are independent, the total
 multiplicative systematic uncertainty $\sigma_{\text{multi}}$  is calculated by summing 
 the individual contributions in quadrature. 
 The results are summarized in Table~\ref{sys}. 

The systematic uncertainty in signal extraction stems from
 the signal shape, background shape, and fit range. The signal shape uncertainty 
 is addressed by varying the width parameter of the convolved Gaussian function within
 1 standard deviation. The uncertainty due to the background shape  
 is evaluated by switching 
 from a second-order to a first-order polynomial function. 
 The uncertainty due to fit range 
   is assessed by adjusting the range by $\pm$10 MeV.
The largest signal yield, accounting for these combined effects, 
is adopted as the final result.
 
 To take into account the multiplicative systematic uncertainty,
 the likelihood curve is further convolved by a Gaussian with
 a width parameter equal to the total multiplicative systematic uncertainty 
 $\int_{0}^{+\infty} \mathcal{L}(N^{\prime}_{\text{sig}}) 
 e^{-\frac{(N_{\text{sig}} - N^{\prime}_{\text{sig}})^2}{2(\sigma_{\text{multi}} N_{\text{sig}})^2}} dN^{\prime}_{\text{sig}}$, 
 where $N^{(\prime)}_{\text{sig}}$ 
  is the signal yield and $\mathcal{L}(N^{\prime}_{\text{sig}})$   
 is the likelihood function. The final result is based on the largest upper limit estimate, with all systematic uncertainties taken into account.


 The systematic uncertainties for $\mathcal{R}_1$ and $\mathcal{R}_2$ 
 align with those
 of the cross section measurement. However, since it is a ratio measurement, several uncertainties cancel, including those from integrated
 luminosity, photon detection, tracking efficiency, kinematic fit, and radiative
 correction. The remaining uncertainties for the search channels $\x\to\Ks\kpm\pi^{\mp}$
and $\x\to K^*(892)\bar{K}$ arise from the PID, MC decay model,
 $\Ks$ reconstruction, and 
 $\pi^0$ veto, totaling 5.4$\%$ and 7.0$\%$, respectively. 
For the normalization channel $\x\to\pi^{+}\pi^{-} J/\psi$,  remaining uncertainties 
stem from the statistical uncertainty of $N_{\pip\pim\jpsi}$, the branching
fraction $\mathcal{B}(\jpsi\to l^{+} l^{-})$, and the $\jpsi$ mass window, totaling 12.4$\%$~\cite{BESIII:2023eeb, BESIII:2023xta}. 
 The total systematic uncertainties for the ratio measurements are 13.5$\%$
 for $\mathcal{R}_1$ and 14.2$\%$ for $\mathcal{R}_2$.

\section{SUMMARY}
In this paper, we investigate the decays $\x\to\Ks\kpm\pi^{\mp}$ and $\x\to K^*(892)\bar{K}$, where the $\x$ is produced through the process $\ee\to\gamma\x$, utilizing
data collected by the BESIII detector at c.m. energies ranging from 4.16 to 4.34~GeV.
No significant signals are observed for either decay mode. As a result, we 
estimate upper limits at the 90$\%$ CL on the product
 of the production cross section and branching 
fractions, denoted as
$\sigma[\ee\to\gamma\x]\cdot\mathcal{B}[\x\to\Ks\kpm\pi^{\mp}]$ and 
$\sigma[\ee\to\gamma\x]\cdot\mathcal{B}[\x\to K^*(892)\bar{K}]$, across the
 specified energy range.

Furthermore, we evaluate the relative branching fractions 
of these decay channels 
compared to the well-established decay $\x\to\pip\pim\jpsi$. These ratios are 
determined as
$\mathcal{R}_1=\frac{\mathcal{B}(\x\to\Ks K^{\pm}\pi^{\mp})}
{\mathcal{B}(\x\to\pip\pim\jpsi)}<0.07$ and 
$\mathcal{R}_2=\frac{\mathcal{B}(\x\to K^*(892)\bar{K})}{\mathcal{B}(\x\to\pip\pim\jpsi)}<0.10$
 at the 90$\%$ CL 
For the decay $\x\to K^*(892)\bar{K}$, using the branching fraction $\mathcal{B}(\x\to\pip\pim\jpsi)=(4.3\pm1.4)\%$
from the PDG~\cite{2024pdg}, we derive an upper limit on the branching fraction of 
$\mathcal{B}[\x\to K^*(892)\bar{K}] <0.60\times 10^{-2}$. Our 
 upper limit falls within the lower region of the theoretical prediction ranges, providing significant constraints on the 
 intermediate  charmed meson loops model parameter $\alpha$~\cite{kst_theo}.
This is not only of significant importance to the theoretical calculations of 
 $\x\to VP$ and $\x\to VV$, but also provides insights into the nature of the $\x$.
In the context of the conventional charmonium scenario, the branching fractions for $\chi_{c1}(1P)\to K_S^0K^{\pm}\pi^{\mp}$ and $K^*K$~\cite{2024pdg} are of the same order of magnitude as the upper limits we calculated for the $\x$ decays into these final states. These comparable values seem not to contradict the interpretation of the $\x$ as a conventional charmonium state.

\textbf{Acknowledgement}

The BESIII Collaboration thanks the staff of BEPCII (https://cstr.cn/31109.02.BEPC) and the IHEP computing center for their strong support. This work is supported in part by National Key R\&D Program of China under Contracts Nos. 2025YFA1613900, 2023YFA1606000, 2023YFA1606704; National Natural Science Foundation of China (NSFC) under Contracts Nos. 11635010, 11935015, 11935016, 11935018, 12025502, 12035009, 12035013, 12061131003, 12192260, 12192261, 12192262, 12192263, 12192264, 12192265, 12221005, 12225509, 12235017, 12361141819; the Chinese Academy of Sciences (CAS) Large-Scale Scientific Facility Program; the Strategic Priority Research Program of Chinese Academy of Sciences under Contract No. XDA0480600; CAS under Contract No. YSBR-101; 100 Talents Program of CAS; Project ZR2022JQ02 supported by Shandong Provincial Natural Science Foundation; The Institute of Nuclear and Particle Physics (INPAC) and Shanghai Key Laboratory for Particle Physics and Cosmology; ERC under Contract No. 758462; German Research Foundation DFG under Contract No. FOR5327; Istituto Nazionale di Fisica Nucleare, Italy; Knut and Alice Wallenberg Foundation under Contracts Nos. 2021.0174, 2021.0299; Ministry of Development of Turkey under Contract No. DPT2006K-120470; National Research Foundation of Korea under Contract No. NRF-2022R1A2C1092335; National Science and Technology fund of Mongolia; Polish National Science Centre under Contract No. 2024/53/B/ST2/00975; STFC (United Kingdom); Swedish Research Council under Contract No. 2019.04595; and the U. S. Department of Energy under Contract No. DE-FG02-05ER41374.


\end{document}